# Human-Centered Tools for Coping with Imperfect Algorithms During Medical Decision-Making

Carrie J. Cai, Emily Reif, Narayan Hegde, Jason Hipp, Been Kim, Daniel Smilkov,
Martin Wattenberg, Fernanda Viegas, Greg S. Corrado, Martin C. Stumpe, Michael Terry
Google Brain, Google Health
Mountain View, CA
{cjcai,ereif,hegde,hipp,beenkim,smilkov,wattenberg,viegas,gcorrado,mstumpe,michaelterry}@google.com

## ABSTRACT

Machine learning (ML) is increasingly being used in image retrieval systems for medical decision making. One application of ML is to retrieve visually similar medical images from past patients (e.g. tissue from biopsies) to reference when making a medical decision with a new patient. However, no algorithm can perfectly capture an expert's ideal notion of similarity for every case: an image that is algorithmically determined to be similar may not be medically relevant to a doctor's specific diagnostic needs. In this paper, we identified the needs of pathologists when searching for similar images retrieved using a deep learning algorithm, and developed tools that empower users to cope with the search algorithm on-the-fly, communicating what types of similarity are most important at different moments in time. In two evaluations with pathologists, we found that these refinement tools increased the diagnostic utility of images found and increased user trust in the algorithm. The tools were preferred over a traditional interface, without a loss in diagnostic accuracy. We also observed that users adopted new strategies when using refinement tools, re-purposing them to test and understand the underlying algorithm and to disambiguate ML errors from their own errors. Taken together, these findings inform future human-ML collaborative systems for expert decision-making.

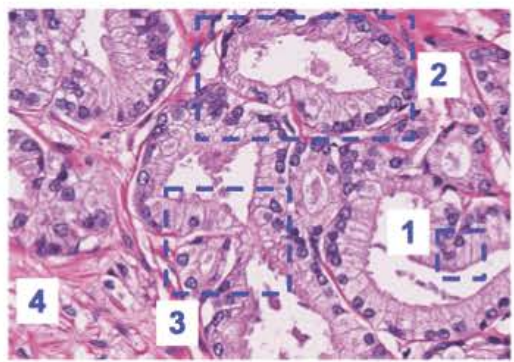

Figure 1: Medical images contain a wide range of clinical features, such as cellular (1) and glandular morphology (2), interaction between components (3), processing artifacts (4), and many more. It can be difficult for a similar-image search algorithm to perfectly capture an expert's notion of similarity, because what's diagnostically important and not important differs from case to case and is highly context-dependent.

## 1 INTRODUCTION

Rapid advances in machine learning (ML) have made it increasingly applicable to problems in the medical domain. One promising application of ML in this space has been in deep neural network (DNN)-backed content-based image retrieval (CBIR) systems. These systems index and retrieve images based on automatically learned similarity metrics [45]. In medical uses of CBIR, doctors use a medical image (e.g., tissue from a biopsy, or an x-ray) as a query for retrieving similar images from previously diagnosed patients [1, 34, 35, 42]. Because similar patients give doctors points of comparison, CBIR has been used to improve medical decision making by filling knowledge gaps, aiding consistency, and refreshing knowledge of rare cases [23]. In this paper, we focus on the needs of pathologists using a DNN-backed CBIR system for making a *differential diagnosis*: the process of differentiating a disease from others that share similar clinical features [8].

During diagnostic decision-making, poor search results from a CBIR system can lead to a brittle user experience. Ideally, a user makes a single query and the search algorithm returns results that address their information needs. However, no algorithm will be able to perfectly capture an expert's notion of ideal similarity for *every* query, given that needs will vary from case-to-case and user-to-user [1, 43]. For example, in pathology, there are numerous visual features of a medical image that could be important at different moments in time, ranging from glandular and cellular morphology to tissue architectures and histologies [12] (Figure 1). Thus, an image that is algorithmically determined to be similar may not be clinically relevant to the pathologist's in-the-moment diagnostic needs. In these cases—when a system fails to return clinically-relevant results—physicians can quickly lose trust in the system and abandon it in favor of their own domain expertise, even if it provides value in other cases [24, 28, 46]. The black-box, opaque nature of ML-based systems can further exacerbate these issues of trust and worsen the overall interactive experience [22, 24, 27]. Interactive mechanisms can help address these issues, by granting the end-user more agency in guiding the search algorithm [43, 47].

In this paper, we propose and evaluate interactive refinement techniques for pathologists to indicate what characteristics of a query image are important when searching for similar images, empowering them to cope with limitations in the search algorithm on-the-fly. While considerable work has focused on improving the accuracy of CBIR search algorithms, comparatively less work has focused on interactive refinement techniques: the types of capabilities users want, how they use them, and how they affect the user experience. This paper contributes to this broader literature by examining what pathologists need when using ML-powered image search, the practices they adopt while using search refinement tools, and the ways in which these refinement tools affect end-user attitudes towards the underlying search algorithm. Though this work focuses on pathologists, our results are relevant to other applications of CBIR that leverage similar image search.

To ground this research, we developed SMILY (Similar Medical Images Like Yours), a prototype application that uses a DNN to identify visually similar medical images. SMILY includes a range of end-user tools to guide the algorithm's retrieval process (Figure 2): 1) *refine-by-region*, which allows users to select a region of interest to focus on in the query image, 2) *refine-by-example*, where users can mark useful search results to retrieve more results like them, and 3) *refine-by-concept*, a novel refinement technique that allows users to specify that more or less of a clinical concept should be present in search results. In two evaluations, we found that refinement tools not only enhanced the utility of information found and increased user trust, but also served purposes beyond providing algorithmic feedback, such as helping users develop a mental model of the ML algorithm, or probe the likelihood of a diagnosis.

In sum, this paper makes the following contributions:

- We enumerate key needs of pathologists when searching for similar images during medical decision-making.
- We present the design and implementation of interactive refinement tools, including a novel technique, *refine-by-concept*, that leverages key affordances of deep neural network models for similarity search.
- We report results from two studies demonstrating that these refinement tools can increase the utility of clinical information found and increase user trust in the algorithm, without a loss in diagnostic accuracy. Overall, experts preferred SMILY over a traditional interface, and indicated they would be more likely to use it in clinical practice.
- We identify ways that experts used refinement tools for purposes beyond refining their searches, including testing and understanding the underlying search algorithm; investigating the likelihood of a decision hypothesis; and disambiguating ML errors from their own errors.

Collectively, our findings inform the design and research of future human-ML collaborative systems for expert decision-making with images, an area that will likely continue to rise in importance across more domains in the coming years.

## 2 RELATED WORK

This paper draws upon prior work at the intersection of clinical decision support systems, content-based image retrieval, interactive machine learning, and deep neural networks.

**Clinical Decision Support Systems**

Clinical decision support systems (CDSSes) provide clinicians with knowledge to enhance medical decision-making [38]. A wide range of CDSSes exist, from systems that make diagnostic decisions, to those that supply information of potential use to medical decision making [7]. In this paper, our focus is on the latter.

Although studies have shown that CDSSes can reduce human error and improve outcomes [19, 32], one traditional impediment to adoption has been the lack of user acceptance and trust: experts may resist using a system if it does not provide relevant information or capture the nuances of human thinking [24, 28, 46]. Our work demonstrates how interactive refinement tools can directly address some of these issues during medical image search.

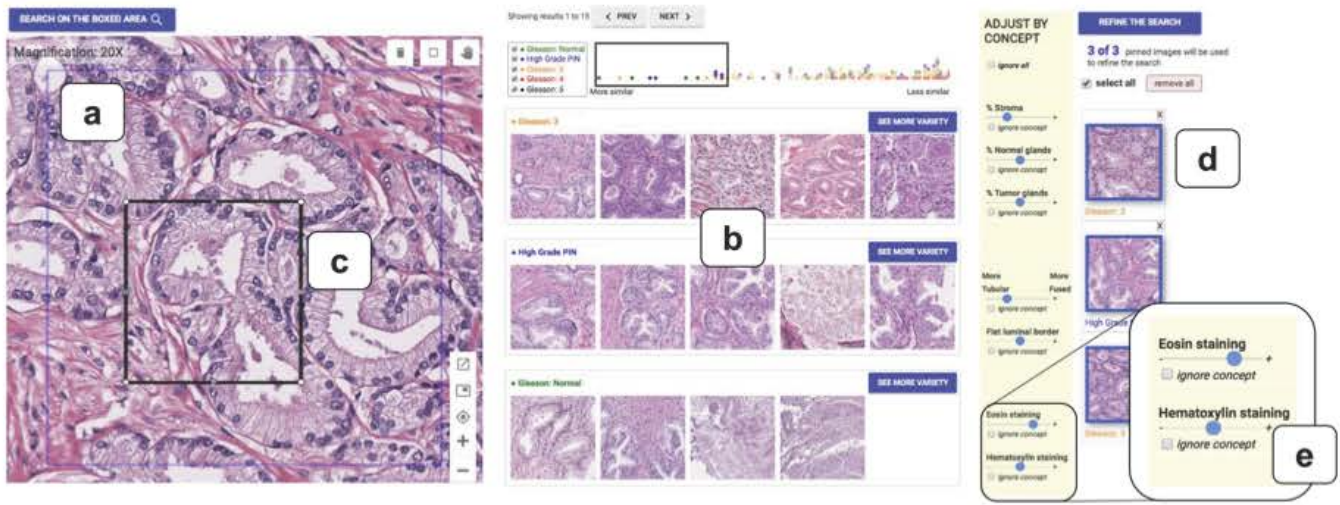

Figure 2: Key components of SMILY: a) the query image (e.g. pathology tissue possibly containing cancer), b) the search results (images from previously diagnosed cases), c) refine-by-region tool: users crop a region to emphasize its importance, d) refine-by-example tool: for clinical concepts that can't be pinpointed to a specific region (e.g. visual patterns), users can pin examples from search results to emphasize that concept, e) refine-by-concept tool: users increase or decrease the presence of clinical concepts by sliding sliders.

### Content-based Image Retrieval

Content-based image retrieval (CBIR) systems allow end-users to search a database of images organized by the images' *content* [11]. Depending on implementation, content can be searched via a query image (i.e., similar image search) or keywords [11, 43]. In medicine, CBIR has been used to help experts make better decisions, by identifying visually similar images for a given query image [1, 23, 34, 35, 42].

A key challenge of CBIR is that the content extracted from an image does not necessarily correspond to the user's semantic interpretation of an image, a phenomenon referred to as the *semantic gap* [43]. To reduce the semantic gap, a wide variety of methods have been proposed to improve the search algorithms, such as automatically extracting color, texture, shape, and other image features [35, 36, 43]. Search results can also be improved via relevance feedback from human annotations, which specify which results are relevant [17, 48]). While a substantial body of work has focused on improving algorithmic capabilities [1, 21, 23, 34, 35, 42], our work examines the user experience surrounding CBIR, an area that has been noted as deserving more attention [11, 35].

### Interactive Machine Learning

Interactive ML incorporates human feedback in the model training process to create better ML models [16], and has been an active area of research [4, 18, 29, 41]. For example, Amershi et al. enabled an algorithm to learn new friend groups on social media given user-provided examples [4], and Fogarty et al. allowed users to interactively teach a search engine new concepts [18]. While much of this prior work has employed handcrafted features [4, 18], we leverage the rich image features (embeddings, described below) automatically learned from deep neural nets. As we will show, these embeddings enable us to introduce interactive refinement mechanisms with relatively little implementation effort, and without modifying the original model.

### Deep Neural Nets (DNN) and Embeddings

Previous work has shown that the features learned in DNNs can be applied to a wide range of tasks, including finding similar images [5, 40]. More specifically, given an input (e.g. image or word), a DNN can produce a corresponding *embedding* [33], or a list of numbers representing the input. This list of numbers can be treated like a coordinate in high-dimensional space, such that items that are more similar are positioned closer together in this space. One key discovery about these embeddings is that the relative locations of embeddings can encode high-level concepts, even if those concepts were not explicitly taught at training time [2, 6, 33]. For example, vector arithmetic on word embeddings can reveal relationships such as *king – queen ≈ man – woman* [33]. Thus, *directions* in an embedding space can encode human-interpretable concepts (e.g. gender) [33]. While different methods have been proposed for computing these directions [6, 13, 33, 49], Kim et al. found that a simple linear classifier can learn these directions quite effectively; those directions are called Concept Activation Vectors (CAVs) [26]. We extend this prior work by showing that CAVs for clinical concepts can be learned from pre-trained DNNs. We expose these CAVs to end users in our refine-by-concept sliders, which allow users to push search results in directions in the embedding space more likely to contain a given concept.

## 3 USER NEEDS

Pathologists study microscopic samples of body fluid or tissue, often to perform cancer diagnosis. To understand what pathologists need during clinical decision-making, as well as what needs arise when they interact with an ML-based CBIR system, we conducted a multi-month iterative design process with 3 pathologists, using mixed methods. We met with pathologists in hour-long sessions approximately every other week. Initially, we used paper prototypes, interviews, and think-alouds to understand pathologists' needs and explore alternative designs. We then created a range of functional prototypes and iterated on those designs with further feedback.

**Needs During Clinical Decision-Making**

When making a differential diagnosis, pathologists need to generate hypotheses, compare and contrast evidence for those hypotheses, and then determine which diagnosis is the most likely. For example, in prostate cancer diagnosis, pathologists compare evidence to assign a grade (called a *Gleason grade* [14]), which indicates the most-likely severity of the cancer. Because cancers in different grades can share similar visual characteristics, this decision can be a difficult one, yet has a pivotal effect on health outcomes: the diagnosis typically determines subsequent patient treatment (e.g. chemotherapy, surgery, or watchful waiting).

In making a differential diagnosis, pathologists first generate a *hypothesis* (e.g. Gleason 3) and a set of *alternative hypotheses* (e.g. Gleason 4, etc.) to rule out. Then, they consider these hypotheses in light of the information they have (e.g., biopsy results, past cases) to determine which is more likely. When they are unsure, they often look for similar images from online tools or textbooks, solicit second opinions from experts, or request further testing. When looking for images similar to their case, pathologists desire to find the most visually similar images across diagnostically distinct categories, to ensure they have not missed a diagnosis. For example, one pathologist explained that *"I have a hypothesis, but I always want that safety net of...what else could this look like?"* Pathologists may iterate on their diagnosis by generating hypotheses and comparing evidence before making a decision.

**Needs Arising from Machine Learning**

To understand needs specific to similar image search, we presented pathologists with paper cut-outs and, later in the process, software implementations of results surfaced by our search algorithm. When interacting with these images, we discovered that pathologists wish to **control which types of similarity matter** for a specific case. More specifically, they wanted to de-emphasize irrelevant features, and emphasize clinically-relevant ones: *"I want to make sure it's triggering on...this gland. I want to say 'that's the thing!'"*

While some features could be localized to a region, other features were pervasive across the entire image, such as overall architectures or visual patterns (our participants used terms such as "sheets," "rosette," and "silky" to describe these features). The relevance of a feature can also change, depending on the pathologist's current focus. Thus, pathologists wished they could push the system to pay attention to different features at different moments in time: *"Maybe you might be interested in the inflammation at one point, but not right now. I would say, no I don't want you to look at the inflammation, I want you to look at everything around it."*

## 4 USER INTERFACE AND SYSTEM DESIGN

Based on identified user needs, we designed and implemented SMILY (Figure 2), a deep-learning based CBIR system that includes a set of refinement mechanisms to guide the search process. Similar to existing medical CBIR systems, SMILY enables pathologists to query the system with an image, and then view the most similar images from past cases along with their prior diagnoses. The pathologist can then compare and contrast those images to the query image, before making a decision. SMILY contains images from a large pathology lab and from pathology image repositories[1].

To find similar images, the query image is fed through a pre-trained deep neural network to retrieve its *image embedding*, a compressed representation of the image that corresponds to a point in high-dimensional coordinate space (as explained in Related Work). Visually similar images are located at points closer together in the embedding space (Figure 3a). SMILY uses a pre-trained, domain-agnostic DNN [21], which we chose given evidence that these deep learning models can achieve performance equivalent to hand-crafted models and can transfer well to a wide range of tasks [5, 45]. Given the query embedding and the pre-computed embeddings of all other images in the database, SMILY finds the most similar images by showing nearest neighbors of the query embedding. Based on user feedback, SMILY displays approximately 15 nearest neighbors per page, to show sufficient variety without overwhelming the user with too many search results.

---
[1]https://tcga-data.nci.nih.gov/docs/publications/tcga/

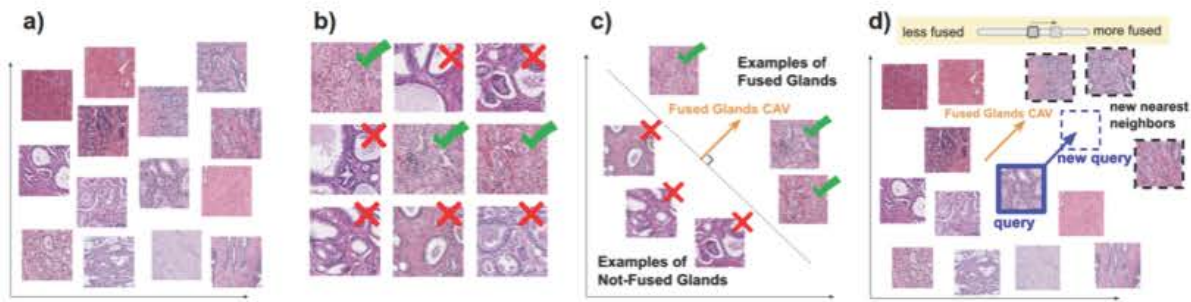

Figure 3: How refine-by-concept sliders are created: a) For each image in the database, SMILY uses a deep neural net to compute its *embedding*, a meaningful representation of the image. Concretely, the embedding is a list of numbers that determines the image's coordinates in a high-dimensional *embedding space*. Although our embedding space has high dimensions (128D), for simplicity we display them here in 2D. b) Users mark a subset of images containing the concept (e.g. "fused glands"). c) A CAV for that concept is learned, corresponding to a direction in the embedding space. d) When a user slides that concept's slider, the original query is shifted in the embedding space in the CAV direction, and new nearest neighbors are found.

**Refinement Tools**

Clinical images can contain a diverse set of visual features, only a subset of which are relevant to the decision at hand (Figure 1). Based on the user needs described above, we created a set of search refinement tools that allow users to emphasize or de-emphasize particular medical concepts when conducting the similar image search. Without altering the deep neural net itself, the tools enable users to push search results towards meaningful directions in the DNN's high-dimensional embedding space.

*Refine-by-region Tool.* Often, the relevant feature can be localized to a physical region of the image (e.g. a specific gland within a biopsy). Using refine-by-region, users can crop a region to isolate a feature, communicating to the system what region is important. While a range of methods could be used (e.g., see [31]), in our implementation, the system simply uses the crop as the new query, and conducts the search only among crops at a similar crop size (e.g., our database contains 300x300 pixel images, as well as 1/4 and 1/8 sized crops of those images). One problem with arbitrary crops is that ML models tend to take inputs with a specific aspect ratio. From early iterations, we found that users desired transparency about which part of the crop would in fact be "seen" by the model. Thus, if a user makes a crop with a very skewed aspect ratio, the system determines the nearest acceptable crop, and displays it in red as an indicator of the true crop that will be fed to the system.

*Refine-by-example Tool.* Cropping alone is sometimes insufficient because the concept of interest cannot be easily isolated to a region. In formative studies, some users pointed to search results containing good examples of concepts, and asked to retrieve more images like those. Based on this observation, and informed by prior work [25, 48], we allow users to adjust the presence of any concept by picking a subset of search results as examples, and updating results on-the-fly to find more images like those examples. To support this type of refinement, we calculate the embeddings of each chosen example, take their average, and use this value as the new search query (ignoring the embedding of the actual query image). This design arose out of formative testing with users, who found results more interpretable when they were strongly anchored to the user-chosen examples.

*Refine-by-concept Tool.* In some scenarios, the clinical concepts of interest may not be present in the returned results, may not be easily isolated within the query image, or may be confounded with other features in the search result images. At other times, it can be helpful to explicitly request that particular medical concepts be present in the returned results, even if they are not present in the query image. This latter capability can be useful to test hypotheses (e.g., *"If this image had more fused glands, how would it affect diagnosis?"*).

To address these needs, we developed *refine-by-concept sliders*, a refinement mechanism that allows an end-user to indicate that they would like more or less of a medical concept to appear in returned results. For example, if the user would like to see search results with more or fewer fused glands, they can adjust the Fused Glands slider to update the search results, emphasizing or de-emphasizing the presence of that concept based on slider settings.

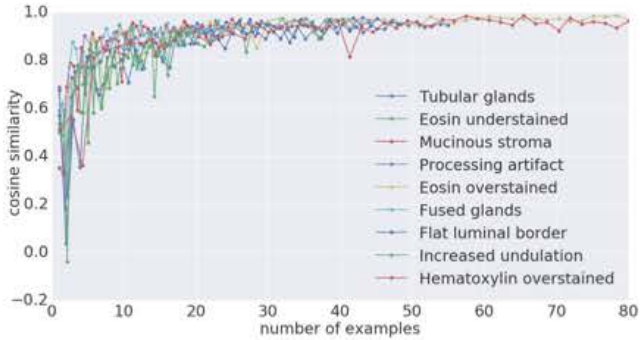

Figure 4: The $y$ axis shows the cosine similarity between CAVs learned on $n$ examples to CAVs learned on all examples for the concept. Overall, CAVs trained on 20 images achieved a high cosine similarity to CAVs learned on all examples for the concept.

Concept sliders leverage image embeddings and directions in the embedding space to affect search results. As noted in Related Work, an image embedding can be thought of as a coordinate in high dimensional space (Figure 3a), and the directions in that space (e.g. man to woman) represent semantically meaningful concepts (e.g. gender). In our implementation, we use simple linear classifiers to learn these directions, referred to as CAVs (described above and in Kim et al. [26]). Our intuition behind using CAVs for creating sliders is as follows: given a learned CAV for a medical concept, adding (or subtracting) the CAV from a query image's embedding effectively pushes the query image's embedding toward (or away from) a concept within the embedding space. When the similarity search is performed with this new, augmented embedding, it will generally return results with more (or less) of the concept.

To identify concepts for this prototype, we asked pathologists to enumerate diagnostically important concepts for prostate cancer diagnosis. Our final set of concept sliders covers a variety of concepts, including structural components, slide-processing artifacts, and visual patterns.

After creating a list of concepts to learn, two pathologists labeled whether the concepts were seen in 100 pathology images (Figure 3b). We then learned CAVs given those labels: for each concept $c$, we use a linear classifier to learn a hyperplane separating the embeddings of examples containing $c$ (positive examples), and embeddings of a random subset not containing $c$ (negative examples). The CAV for that concept ($V_c$) is defined to be the vector orthogonal to the hyperplane (Figure 3c) [26]. Although this process requires some human-labeled data, we empirically evaluated how many labels were needed to produce a reasonable CAV. We found that CAVs trained on 20 labels achieved a high cosine similarity ($\sim 0.9$) with CAVs trained on all labels for that concept (Figure 4).

Finally, we expose each CAV, $V_c$ as a slider in the interface. Moving the slider shifts search results in the direction of the CAV (Figure 3d). For example, given a query embedding $q$, a +0.5 slider shift translates $q$ by $+0.5 * V_c$, yielding new query $q'$. The new search results are produced by taking the nearest neighbors from $q'$.

During formative studies, users were sometimes surprised when search results in the *negative* CAV direction did not reflect the opposing concept. For example, users expected the opposite of Fused Glands to show not-fused glands, but instead it sometimes showed no glands at all. Based on this feedback, we trained *relative* CAVs using an opposing concept as negative examples. Thus, it may be valuable to understand from end-users their expectations for the negative direction of a CAV, such as whether it ought to reflect the absence of a concept versus the opposite concept.

*Auxiliary Refinement Tools.* To support these refinement tools, we implemented several auxiliary tools that help users compare and contrast alternative hypotheses. Here, we briefly summarize those features:

- Narrow hypotheses: While ML algorithms typically output a global list of most similar images, pathologists wished to explicitly control which diagnostic categories to compare and contrast. Thus, users can explicitly filter the images and group by only the diagnostic categories they wish to compare.
- Augment variety: When ruling out alternative hypotheses, users sometimes wanted to see multiple variants within those categories to ensure they had not missed a variant. Thus, users can click to see subgroups of a particular diagnostic category, generated using standard clustering techniques.
- Visualize refinement: A scatterplot overview showing each search results' embedding distances from the query image, color-coded by diagnosis. The visualization changes after every refinement update to show changes in the distribution.

## 5 TOOL EVALUATION STUDY

We evaluated SMILY and its refinement capabilities in two studies. First, we conducted a Tool Evaluation Study (described in this section) to validate that refinement mechanisms update search results in the ways intended. Then, in a User Study (described in the next section), we holistically evaluated how SMILY ultimately affects user experience and search practices during a medical task.

In this section, we consider the former question of whether the refinement mechanisms update results in the ways intended. To this end, we applied each tool to modify the presence of clinical concepts in search results, and asked pathologists to rate the presence of that concept, with and without refinement (in a pairwise manner). We randomized which side the conditions were on, so that raters could not tell which search results were collected with and without refinement.

**Refine-by-region Evaluation**

A common scenario in which a user may want to crop a region is when the algorithm has over-emphasized a physically prominent but irrelevant feature, while under-emphasizing a small, important feature. We collected a random set of ten images where the key glandular structural component was small (approximately 25% of the image). One pathologist cropped the glandular component in the images, after which two other pathologists rated the presence of that structure in search results (on a 7-point scale), with and without refinement.

Images contained a greater presence of the desired feature with refinement ($\mu = 5.3$, $\sigma = 1$) than without ($\mu = 2.8$, $\sigma = 1.4$, ANOVA $p < 0.0005$, $F = 88$). In 88% of cases, with-refinement images contained a greater concept presence. Ratings were tied in 11% of cases. These tended to occur when the original results already contained a relatively high presence of the concept.

**Refine-by-example Evaluation**

The refine-by-example tool can theoretically be used on any concept, not only those that are physically dissectable. We thus selected representative concepts that span the characteristics described above (see System Design), including: Percent Stroma (structural component), Eosin Overstained (artifact), Fused Glands (pattern), and Tubular Glands (pattern). Two morphological pattern concepts were chosen because patterns are most commonly used in decision making.

To collect reference images for evaluation, we consulted with pathologists to enumerate common situations in which they needed to emphasize or de-emphasize each concept. For each concept, a pathologist selected 10 query images meeting those conditions, totaling 40 images. Next, a different pathologist used the tool to refine initial search results, given the desired concept. Finally, two other pathologists rated the presence of the desired concept, blind to condition.

Images contained a greater presence of the desired concept with refinement ($\mu = 4.6$, $\sigma = 1$) than without ($\mu = 2.8$, $\sigma = 1.1$, $p < 0.0005$, $F = 108.7$). With-refinement results contained a greater presence of the concept in 82% of cases, tied in 11%, and had less of the concept in 7%. In the latter, we found that either there were no good examples available in the results, or the chosen examples were "impure," containing a diversity of concepts.

**Refine-by-concept Evaluation**

To evaluate CAVs, we used the same concepts and images from the refine-by-example study. If a CAV reflects the concept well, adding the CAV to a point in the embedding space should lead to a location where images are more likely to contain that concept. Thus, we shifted each query image embedding in the direction of the CAV, and retrieved nearest neighbors. Based on pathologist ratings, images contained a greater presence of the desired concept with refinement ($\mu = 5.4$, $\sigma = 0.7$) than without ($\mu = 2.6$, $\sigma = 1$, $p < 0.0005$, $F = 584.6$). In almost every case (99%), with-refinement results contained a greater presence of the concept, with fewer unintended consequences than refine-by-example.

If a CAV had captured its concept very well, a pathologist ought to be able to infer the concept without being told what the concept is. To investigate this, we asked a third, independent pathologist to identify what concept(s) were being captured by each CAV, by showing the image sets but withholding the concept name (a free-response task). The pathologist correctly identified most concepts without noting other extraneous concepts. However, for Fused Glands, they guessed 3 confounding concepts as well, 2 of which they indicated were biologically correlated with fused glands. The pathologist found the presence of biologically correlated concepts less surprising compared to those that were not biologically correlated with the intended concept. We will discuss these effects in greater detail in later sections.

## 6 USER STUDY

The study reported above confirmed that refinement mechanisms can update search results in intended directions. However, the question remains as to how such tools affect the end-user experience. For instance, refinement tools could hypothetically add complexity and workload to user interactions, possibly decreasing the "seamless" aesthetic [10] of more conventional interfaces (e.g. a simple n-best list); hence, their utility ought to outweigh the costs. More importantly, it is unknown how refinement mechanisms are actually used in practice, and their effect on user attitudes. To understand these questions, we compared the experiences of pathologists using SMILY to that of a baseline conventional interface during prostate cancer assessment. We based the conventional interface on traditional medical CBIR systems [23, 35]. Like traditional medical CBIR systems, the conventional interface displays an n-best list of most similar images, ranked in order of similarity, arranged in a grid. The questions our study sought to answer were:

(1) Does SMILY increase the utility of information for diagnostic decision-making? How does it affect workload and trust, compared to a conventional n-best list interface?

(2) How do pathologists use refinement tools in their search and decision-making practices? What are the trade-offs between different kinds of refinement tools?

**Measures**

We evaluated the following outcome metrics related to utility for decision-making, workload, and attitudes towards the system. All items below were rated on a 7-point scale.

- Diagnostic utility: Participants answered the question *"How useful were the examples you found in assisting with diagnosis?"*
- Mental support for decision-making: Participants answered the Likert scale question *"[Version X] helped me think through the diagnosis and organize my thoughts."*
- Workload: Participants answered the effort and frustration dimensions of the NASA-TLX [20].
- Trust: There exist a variety of approaches and questionnaires for measuring trust. In this study, we used Mayer's dimensions of trust [30] because these have been widely used in prior studies on trust and referenced in existing HCI work. Participants answered Likert scale questions on the system's capability and benevolence, key dimensions of trust.
- Future use: Participants answered the Likert scale question *"I would continue using [the system] in my practice."*
- Overall preference between the two interfaces: Participants rated on a 7-point Likert scale ranging from 1 (totally version A), 2 (much more version A than B), 3 (slightly more version A than B), 4 (neutral), etc. to 7 (totally version B).

In the study, we called the two systems "Version A" and "Version B" (counterbalanced) to avoid biasing participants, but we refer to them here as "SMILY" and "conventional interface" for clarity.

**Method**

Twelve pathologists participated in the user study. All had 1-20 years ($\mu = 9.8$) of pathology experience post-residency training. Each participant first completed an online tutorial of both interfaces on their own (30 minutes), then analyzed 6 prostate images (1.5 hours total), followed by a post-study questionnaire and semi-structured interview (30 min).

To ensure the task was not trivial, we first identified images that had previously been contested between pathologists, with conflicting diagnosis labels. Then, to cover a range of scenarios, we identified 6 cases spanning 3 image types: 2 "borderline" images (classically difficult cases lying on the border between cancer grades), 2 "asymmetric" images (images in which an irrelevant feature was larger than a relevant feature), and 2 "diverse" images (images containing a diversity of clinical patterns). Participants used SMILY for 3 trials

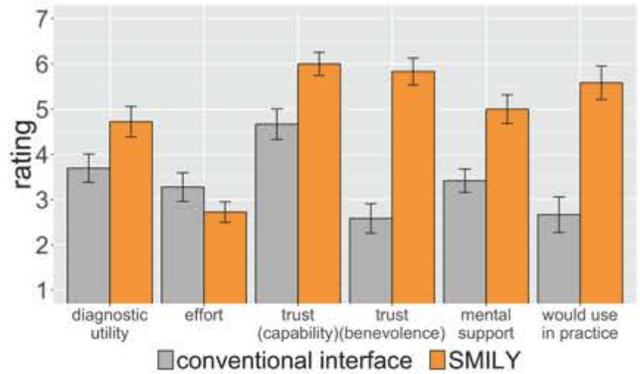

Figure 5: Results from user study survey questions, with standard error bars. Compared to a conventional interface, SMILY had greater diagnostic utility and required lower effort. Users also indicated having greater trust in SMILY, felt SMILY offered better mental support, and felt they were more likely to use it in clinical practice.

(1 image in each category), and the conventional version for 3 trials. Order of conditions and categories were randomized.

After each trial, pathologists answered the diagnostic utility, effort, and frustration questions, as well as the diagnosis categories they saw in the image, along with their decision certainty. During each trial, participants' actions were recorded using a log file and by screen recording. In the post-study questionnaire, participants answered the remaining questionnaire items.

## 7 USER STUDY RESULTS

Figure 5 summarizes the ratings from participants. During the study, participants found the clinical information to have **higher diagnostic utility** while using SMILY ($\mu = 4.7$) than while using the conventional interface ($\mu = 3.7$). A mixed-effects regression, with interface and image type as fixed effects and participant as a random effect, found a significant main effect of interface on usefulness ($p = 0.025$, $t = 2.3$).

Participants also experienced **less effort** using SMILY ($\mu = 2.8$) than the conventional interface ($\mu = 3.3$). Results show a significant main effect of interface on effort ($p = 0.034$, $t = -2.2$), as well as a main effect of image type ($p = 0.006$, $t = -2.9$, diverse image type > borderline image type). No differences were found for the frustration dimension.

In the post-study questionnaire, users expressed **higher trust** in SMILY. Users rated SMILY as having higher capability than the conventional interface ($\mu = 6$, $\mu = 4.7$, $p = 0.01$, $t = 3.08$), as well as higher benevolence ($\mu = 5.8$, $\mu = 2.6$, $p < 0.001$, $t = 6.04$).

Participants found SMILY to offer **greater mental support** for decision-making than the conventional interface ($\mu = 5$, $\mu = 3.4$, $p = 0.003$, $t = 3.8$), and expressed that they would be be **more likely to use** SMILY in clinical practice ($\mu = 5.58, \mu = 2.67, p = 0.001, t = 4.4$). Overall, 7/12 participants "totally" preferred SMILY, 4/12 preferred SMILY "much more" than the conventional interface, and 1/12 preferred SMILY "slightly more." Users felt that, compared to the conventional interface, SMILY would be more practically useful in their day to day work: *"That is where I would do the lion's share, 85% of my work."*

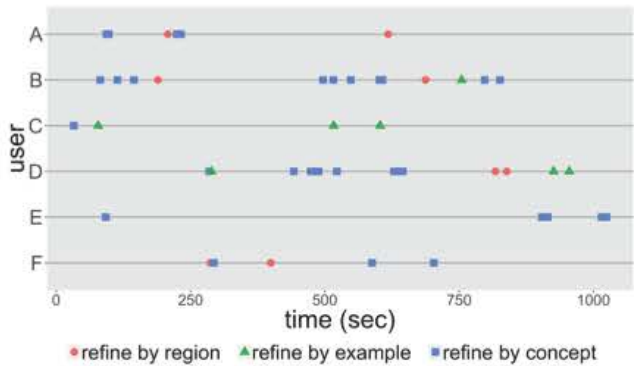

Figure 6: Participants interleaved between different refinement tools, using a variety of strategies to emphasize and de-emphasize clinical features. The figure shows how different participants used refinement tools for the same prostate cancer case. Thus, data from half (6) the participants are shown due to counterbalanced conditions.

In sum, users were able to retrieve more diagnostically relevant information with less effort using SMILY, had greater trust in the system, and felt they were more likely to use it in clinical practice. While the focus of this work is on the refinement mechanisms as opposed to diagnostic accuracy, SMILY ideally should not dampen accuracy. We found no significant effect of interface on accuracy (p=0.55) and confidence (p=0.7). However, the detection of subtle differences would require a full clinical trial, a direction that is promising but outside the scope of the current study.

## 8 TOOL USE AND NAVIGATION PATTERNS

In this section, we describe how individual refinement tools and the conventional interface were used to locate desired images. We discuss support for decision making in the next section.

### Refine-by-Region

The refine-by-region is the only tool that operates directly on the image in question. Participants used it to search on specific regions of interest, and to exclude other intertwined features from being considered by the algorithm. Hence, refine-by-region tended to initiate mini-search sessions dedicated to a specific area. A common pattern was to make coarse-grained refinements using refine-by-region, then finer-grained adjustments using refine-by-example and refine-by-concept.

### Refine-by-Example

Pathologists used refine-by-example when they instinctively saw images containing features they cared about: *"It was...an intuitive feeling of 'this looks like this.'"* Although refine-by-example was easy to initiate due to its perceptual nature, users eventually encountered friction while using the tool. When changes in search results were subtle, users were sometimes uncertain about whether the results were improving. This was more common when the examples they picked contained many confounding features, which further muddied the results. In analyzing user logs, we found that users took a relatively long time on each iteration of refine-by-example (median=86 sec), with these chains of iterations consuming a median of 139 seconds. Users also tended to provide only a few examples per update ($\mu = 2, \sigma = 1$). In contrast, the refine-by-concept tool had faster iterations (median=15 sec), with chains of consecutive iterations taking a median of 91 seconds. The slower iterations during refine-by-example were likely due to the users' time spent picking examples and, in cases where changes were subtle, determining whether search results were updating in the directions intended.

### Refine-by-Concept

Refine-by-concept and refine-by-example played complementary roles, and appeared to trigger different types of thinking. Whereas interaction with refine-by-example was primarily perceptual and pattern-based (*"You think in pictures first before you think in words..."*), refine-by-concept pushed users to think more systematically about the specific features important for diagnosis: *"The [refine-by] concept makes you think a little more about what components make the image look like what it does."*

Unlike other tools, refine-by-concept enabled users to explore the *spectrum* of a key differential feature using sliders. For example, because the extent of gland fusion is a key factor that determines whether a specimen is grade 3 versus grade 4 cancer, some used the Fused Glands slider to view a range of fused and un-fused glands, along with their previous diagnoses, to determine where their new case stood on that spectrum: *"If there is a question between 2 grades, you can shift them in so many ways and see a lot of different possibilities."* Users emphasized that the distinction between two diagnoses is often continuous rather than discrete, making the sliders particularly useful for borderline cases: *"The zone between 3 and 4, it's hard to make an exact black and white line, so it was helpful for that differential."*

Refine-by-concept was also helpful in situations where search results were "stuck" in a suboptimal portion of the embedding space, making it difficult to find better examples to refine on. For example, a user realized that the examples he marked for refine-by-example had the same irrelevant feature as the original reference image: *"I don't think it's getting any better...partly [because] I'm picking ones that have a lot of stroma."* He overcame this problem by using refine-by-concept: *"Yeah so that helps a lot! So the key there was not to refine by example but to adjust by concept."*

Although rare, refine-by-concept sliders sometimes surfaced results containing an unrelated concept in addition to the target concept, due to a confounding visual variable. For example, placing Fused Glands on the maximum slider setting returned some images of benign stroma, which have small dots that are visually similar to hyper-fused glands. While confounding concepts can be interpretable if they are biologically correlated with the intended concept, in this case, it was jarring because the visually similar concepts conflicted biologically. To help control for the presence of confounding variables, future work could investigate the effects of including visual mimickers of the target concept as negative examples in the training of CAVs.

In sum, users found refine-by-concept very useful, and many asked for more concepts to be made available. For example, users suggested concepts such as "sheets of cells," "prominence of nucleoli," "mitotic activity," and "benign mimickers of cancer." Although some human-labeled data was needed, we found that the number of labels required was relatively small (see System Design), giving hope that this technique can be applied to more concepts and domains. For example, future work could explore enabling end-users to train concepts on-the-fly. Research could also tackle how to learn concepts that are less common in medical datasets, such as by using examples from textbooks, journals, or even image search, to surface enough examples to learn the concept. Overall, making dynamic creation of any concept scalable and accessible will be important future work.

### Interleaving, Resets, and Backtracks

Within a single case, users often interleaved between tools multiple times (Figure 6), sometimes to backtrack and switch to a different refinement strategy, and sometimes to fine-tune progress made in previous steps. In refine-by-example, a common pattern was to "clear the slate" by deleting all image examples at once before adding more. These behaviors tended to occur after having used a different tool in between, suggesting that the search results may have shifted enough to provide better examples for selection. Within a single tool, users also backtracked, for example, with refine-by-concept sliders: *"It went a little too far... Let's see if I can back off a little."*

### Navigation in the Conventional Interface

While refinement enabled users to narrow in on the most relevant dimensions of similarity, in the conventional interface users resorted to sequentially scanning through pictures, skipping over ones that were irrelevant to the diagnosis. Users described it as *"just kind of flipping through the book"*, *"keep going next"*, and *"looking through possibly hundreds of images, which is very time consuming."* Activity logs corroborated these findings: users went through multiple pages of results to find relevant ones ($\mu = 2.6$), whereas with SMILY they usually stayed on the first page and refined the search ($\mu = 0.4$). With the conventional interface, users wished they had more control over the search direction: *"I'm lamenting the fact that I can't do more refine-by-example and I'm kind of stuck going through all of these things."*

## 9 DECISION MAKING AND COPING WITH BLACK-BOX ML

The process of clinical decision-making involves generating hypotheses, comparing and contrasting them, and determining their likelihood. In this section, we first describe the refinement strategies used for navigating the decision-making process. Then, we describe refinement strategies used for coping with the ML algorithm.

### Refinement Practices for Decision-making

*Tracking the Likelihood of a Decision Hypothesis.* During the study, users described ways in which iteratively refining search results helped them track the likelihood of a hypothesis. Some commented that, with each round of refinement, seeing an increasing number of visually similar images appear in their hypothesized category gave them reassurance that they were on the right track: *"When you can refine by example you can see hopefully more of the same...and feeling more assurance that that's the right answer."* Pathologists also used iterative updates as an important signal for determining if they were on the wrong path: *"It's kind of like looking at an object from a bunch of different views. That will either increase your confidence...or it will show you that you're walking down the wrong path."*

*Generating New Ideas.* The refinement tools also provided decision-making support by helping users reflect on their thought process and generate new ideas: *"The iterative changes help you think through what did I do, how did I get there."* In particular, the concept sliders helped raised awareness of unexplored territories: *"If I glance at the concept [slider], it may make me think 'Oh but I didn't try to do the luminal flattening,' whereas that would not have come to mind on its own."* One user remarked that a potential downside of this is the possibility of going too deep into a thought process: *"The [refinement tools] helped me think a lot, maybe too much."* As the possible range of diseases can be large, pathologists may be wary of going too deep down a tangential rabbit hole, if it takes them too far from their initial thought process.

*Reducing Problem Complexity.* Reference images containing a mixture of features were often cognitively difficult to analyze. As found in the results above, users experienced greater effort when analyzing images with diverse features, perhaps because they required analyzing each feature separately while simultaneously considering their interactions: *"These are different patterns. I can't do it all at once."* The crop tool helped users focus on one component at a time, potentially offloading intermediate state from working memory to the user interface.

**Refinement Strategies for Coping with ML**

Interaction with a search algorithm during the decision-making process introduced additional challenges and uncertainty. In this section, we discuss ways in which users attempted to resolve these challenges, and how they made use of refinement tools in the process.

*Attempting to Narrow the Semantic Gap.* Users were surprised when their mental model of similarity did not match that of SMILY, especially when the system missed a key, diagnostically-critical feature or over-emphasized obviously irrelevant features. Users found irrelevant images distracting: *"It's sort of a red herring, because yes it looks like this image, but it's not important."* This semantic gap degraded trust, particularly in cases where users could not fathom the system's reasoning: *"I don't quite trust the system yet, gosh it [the search result] looks nothing like my grade 3 [the reference image]. That fact is making me doubt this thing."* Using refinement tools, participants attempted to narrow the semantic gap, by emphasizing relevant medical concepts and reducing irrelevant ones. By adapting to users' interests [30], refinement tools may have increased SMILY's perceived benevolence and trustworthiness. In contrast, when using the conventional interface, users felt their only option was to look at more pages of search results to get more relevant images, but felt that effort may be futile given search ranking: *"I could click 'next' but that's going down to the less similar ones."*

*Developing a Mental Model of the ML.* Unexpected search results created an additional layer of uncertainty for users, causing them to wonder what the machine was "thinking" and to form theories around these behaviors. Many looked for features in the reference image that could explain why surprising images were being returned: *"I wonder what it was looking at...must have been picking up on some of these single nuclei."* Some formed elaborate theories about what ratios of features the algorithm had paid attention to (*"This is a quarter of the [image] and it's got a big benign gland...sticking in"*), or developed theories about the goal of the system ( *"I suspect [it's] trying to mimic the workings of the human brain."*) Participants used refinement tools to test and revise their theories: *"To test my hypothesis, I [want to] cut out that second gland."* These observations support growing evidence that interactive seams in an interface could help users manage and refine their algorithmic folk theories [15].

*Disambiguating ML Errors from Self Errors.* Pathologists also used refinement tools to disambiguate ML errors from their own errors. When faced with surprising search results, users wondered whether the algorithm was simply noisy, or whether it had seen something important that they themselves had missed: *"I don't know why it's picking up PIN [Prostatic Intraepithelial Neoplasia] so much! Now I'm questioning myself"* or *"They make me wonder, 'Oh, am I making an error?' "* This need for disambiguation is critical to decision-making: if the algorithm is behaving properly, then unexpected results could be fortuitous as they cause experts to stop and question their own hypotheses. However, if the algorithm is wrong, this could add additional burden on the practitioner. To test if the error was due to ML, pathologists removed variables they thought were leading the algorithm astray and repeated the search. Over time, some tried to remove these variables *preventatively*, by cropping out areas known to introduce noise. Thus, without providing an explicit explanation about how the algorithm works, SMILY implicitly gave users some insight by allowing them to test the algorithm.

*Fear of Over-influencing the Algorithm.* In some cases, setting the More Fused slider to its maximum setting displayed examples of items that were beyond fused. One user was concerned that the powerful nature of the sliders might over-influence the algorithm: *"If I'm adjusting that bar, is there a way to say 'keep your top selection, you were giving me these pattern 4s'? If I'm injecting too much of my interpretation into it, how much of this is [me] putting in my subjective interpretation hoping to get that response back?"* In sum, while users desired to control the *visual* characteristics, they wished the *diagnostic* categories could be preserved.

## 10 DISCUSSION

In light of our findings, we now discuss the broader implications of this research.

**Applicability of Refinement Tools**

As machine learning algorithms continue to improve in accuracy and reduce errors, some might question the continued value of refinement tools. First, as explained earlier, it may be impossible to develop an algorithm that perfectly captures an expert's notion of similarity, given that these notions differ by case, and often even within a single case as the user moves from one focus area to another. In domains where data is highly regulated and expert time is rare, it can be exceedingly cumbersome to obtain new expert-labeled data sets every time a model needs to be improved. Our findings show that lightweight refinement tools enabled users to more easily customize a system for their in-the-moment needs, without re-training the original model. While we demonstrated efficacy in pathology, the methods themselves were not specific to one domain. This gives us hope that the techniques we have developed can be applied to other sub-domains where the user task is to make a critical decision using image data.

Second, we saw that refinement tools were used for purposes beyond improving search results, such as iteratively tracking a diagnosis and building mental models of the ML, leading to greater trust. As seen in the past, medical experts are likely to resist automation if it appears brittle or limits their autonomy [24, 46]. Our results support the notion that automation and agency are not necessarily trade-offs, but are instead mutually beneficial. When faced with an imperfect algorithm, refinement tools gave experts the agency to guide the system. This in turn allowed the system to improve results and demonstrate partial capability, increasing trust. As artificial intelligence continues to play a more prominent role in domains traditionally held by human experts, we view these tools as being even more crucial to the usability and utility of ML.

### Medical Decision-Making and Bias

Relative to other search domains, CBIR systems for decision-making face a unique challenge because both humans and algorithms operate under decision uncertainty: when unexpected system behavior occurs, the user may not know whether it was due to an algorithmic error or their own oversight. In SMILY, refinement tools gave experts a way to start disambiguating between these causes. Designing end-user tools for disambiguation will be critical to the future effectiveness of these systems.

While refinement tools helped users test for ML errors, a potential risk of refinement is confirmation bias. Since experts used iterative refinement as a means for tracking the likelihood of a hypotheses, this could lead to bias if users are searching only towards evidence consistent with their existing beliefs [37], or believe results are improving when they're not [44]. Though we did not find evidence of deteriorated decision-making using SMILY, the potential for confirmation bias should be considered in future work. For example, systems could mitigate confirmation bias by highlighting refinement paths that the user has *not* yet taken, or by raising awareness when a refinement update has increased evidence in a diagnostic category outside of the user's current focus.

### Beyond Algorithmic Feedback: Refinement as a Means for Testing and Understanding Opaque Algorithms

While human-ML interactive tools have traditionally been used to improve algorithms, we found that refinement mechanisms empowered humans to test, understand, and grapple with opaque algorithms. These findings suggest new ways for improving algorithmic transparency, which to date has focused more on generating human-interpretable algorithmic explanations [9, 26, 39]. Beyond being passive recipients of machine output, end-users could play an active role in the interpretation of machines, equipped with interactive tools to hypothesis-test their intuitions. Indeed, interaction could help people form mental models and increase algorithmic transparency [3, 15]. Integrating these human-centric approaches with existing efforts is a promising direction for future research, as it opens up possibilities for leveraging the intelligence of human beings themselves.

## 11 CONCLUSION

In this paper, we found that refinement tools not only increased trust and utility, but were also used for critical decision-making purposes beyond guiding an algorithm. Our work brings to light the dual challenges and opportunities of ML: although black-box ML algorithms can be difficult to understand, off-the-shelf image embeddings from DNNs could enable new, lightweight ways of creating interactive refinement and exploration mechanisms. Ultimately, refinement tools gave doctors the agency to hypothesis-test and apply their domain knowledge, while simultaneously leveraging the benefits of automation. Taken together, this work provides implications for how ML-based systems can augment, rather than replace, expert intelligence during critical decision-making, an area that will likely continue to rise in importance in the coming years.

## 12 ACKNOWLEDGEMENTS

We are grateful to the following individuals for their valuable help and feedback on this work: Trissia Brown, Michael Emmert-Buck, Isabelle Flament, Fraser Tan, Lily Peng, Craig Mermel, Mahul Amin, Niels Olson, Mahima Pushkarna, Jimbo Wilson, Andrei Kapishnikov, Aleksey Boyko, Ed Chi, Ian Li, and Daniel Tse.